\documentclass[numberedappendix,appendixfloats]{emulateapj}
\usepackage{url}

\shorttitle{The IMF at $M$ $\lesssim$ 1~$M_\odot$}
\shortauthors{Kalirai et~al.}

\begin{document}


\title{Ultra-Deep {\it Hubble Space Telescope} Imaging of the Small Magellanic Cloud:  \\
The Initial Mass Function of Stars with $M$ $\lesssim$ 1~$M_\odot$\altaffilmark{1}}


\author{
Jason S.\ Kalirai\altaffilmark{2,3},
Jay Anderson\altaffilmark{2},
Aaron Dotter\altaffilmark{2},
Harvey B.\ Richer\altaffilmark{4},
Gregory G.\ Fahlman\altaffilmark{5}, \\
Brad M.~S.\ Hansen\altaffilmark{6},
Jarrod Hurley\altaffilmark{7},
I.\ Neill Reid\altaffilmark{2},
R.\ Michael Rich\altaffilmark{6}, \\
and Michael M.\ Shara\altaffilmark{8},
}
\altaffiltext{1}{Based on observations with the NASA/ESA {\it Hubble Space 
Telescope}, obtained at the Space Telescope Science Institute, which is operated 
by the Association of Universities for Research in Astronomy, Inc., under NASA 
contract NAS5-26555.  These observations are associated with proposal GO-11677.}
\altaffiltext{2}{Space Telescope Science Institute, 3700 San Martin Drive, Baltimore, 
MD, 21218; jkalirai@stsci.edu; jayander,dotter@stsci.edu}
\altaffiltext{3}{Center for Astrophysical Sciences, Johns Hopkins University, Baltimore, MD 21218, USA}
\altaffiltext{4}{Department of Physics \& Astronomy, University of British Columbia, 
Vancouver, BC, Canada; richer@astro.ubc.ca}
\altaffiltext{5}{National Research Council, Herzberg Institute of Astrophysics, Victoria, BC, 
Canada; greg.fahlman@nrc-cnrc.gc.ca}
\altaffiltext{6}{Division of Astronomy and Astrophysics, University of California at Los Angeles, 
Los Angeles, CA, 90095; hansen/rmr@astro.ucla.edu}
\altaffiltext{7}{Center for Astrophysics \& Supercomputing, Swinburne University of Technology, 
Hawthorn VIC 3122, Australia; jhurley@swin.edu.au}
\altaffiltext{8}{Department of Astrophysics, American Museum of Natural History, Central Park 
West and 79th Street, New York, NY 10024; mshara@amnh.org}


\begin{abstract}

We present a new measurement of the stellar initial mass function (IMF) based on ultra-deep, 
high-resolution photometry of $>$5,000 stars in the outskirts of the Small Magellanic 
Cloud (SMC) galaxy.  The {\it Hubble Space Telescope (HST)} Advanced Camera for Surveys (ACS) 
observations reveal this rich, co-spatial population behind the foreground globular cluster 47~Tuc, 
which we targeted for 121 {\it HST} orbits.  The stellar main sequence of the SMC is measured in the 
$F606W$, $F814W$ color-magnitude diagram (CMD) down to $\sim$30th magnitude, and is cleanly 
separated from the foreground star cluster population using proper motions.  We simulate the SMC 
population by extracting stellar masses (single and unresolved binaries) from specific IMFs, and 
converting those masses to luminosities in our bandpasses.  The corresponding photometry for these 
simulated stars is drawn directly from a rich cloud of 4~million artificial stars, thereby accounting 
for the real photometric scatter and completeness of the data.  Over a continuous and well populated 
mass range of $M$ = 0.37 -- 0.93~$M_\odot$ (i.e., down to a $\sim$75\% completeness limit at $F606W$ = 28.7), 
we demonstrate that the IMF is well represented by a single power-law form with slope $\alpha$ = $-$1.90 
($^{+0.15}_{-0.10}$) (3$\sigma$ error) (i.e., $dN/dM \propto$ $M^{\alpha}$).  This is shallower than the 
Salpeter slope of $\alpha$ = $-$2.35, which agrees with the observed stellar luminosity function at higher 
masses.  Our results indicate that the IMF does {\it not} turn over to a more shallow power-law form within this 
mass range.  We discuss implications of this result for the theory of star formation, the inferred 
masses of galaxies, and the (lack of a) variation of the IMF with metallicity.

\end{abstract}



\section{Introduction} \label{sec:intro}

Understanding the initial mass function (IMF) of stars is one of the most important and sought after 
pursuits of astrophysics.  The stellar IMF holds its origins in the theory of star formation, and 
is imprinted through physical processes such as turbulence, gravitational fragmentation of clouds, accretion 
in dense cores, and ejection of low mass objects (Larson 1981; Bonnell, Larson, \& Zinnecker 2007).  
The physics of these processes can be constrained by measuring the shape and universality of the IMF 
in vastly different environments, such as metal-rich and dense star forming regions in disks, metal-poor 
and sparse spheroid populations, and gravitationally-bound star clusters.  Additionally, the IMF 
serves as a key input to many interesting problems in astrophysics.  The integral of the function at low 
masses ($\lesssim$1~$M_\odot$) determines 
the Milky Way mass budget including the number of substellar objects, the slope at intermediate 
masses ($\gtrsim$1~$M_\odot$) establishes the level of chemical enrichment into the interstellar 
medium, and the shape of the function at higher masses controls the amount of kinetic feedback 
that stellar populations impart to their surroundings.  Ultimately, the IMF represents a key 
ingredient to general studies of distant galaxies by providing insights on the mapping between 
unresolved light from a mix of stellar populations to fundamental properties (e.g., star formation 
history and mass-to-light ratios).

There is a rich history of astronomical studies aimed at characterizing the IMF, as summarized 
in recent reviews by Chabrier (2003), Bastian, Covey, \& Meyer (2010), and Kroupa et~al.\ (2011).
The majority of previous investigations have calculated the distribution of masses from 
observations of stars near the Sun.  \cite{salpeter55} found that the smoothly-varying 
luminosity function of these stars (from $M_V$ = $-$4 to $+$13) was reasonably approximated 
by a power-law form with (slope) $\alpha$ = $-$2.35 over a mass range of 0.4 -- 10~$M_\odot$ (i.e., 
$dN/dM \propto$ $M^{\alpha}$).\footnote{We adopt the convention of a ``linear'' 
slope for the IMF, where a Salpeter power law has $\alpha$ = $-$2.35.  This is equivalent to 
a logarithmic slope of $\Gamma$ = $-$1.35.}  For $M \gtrsim$ 1~$M_\odot$, this initial work 
has been largely verified by subsequent analysis involving improved local stellar luminosity functions, 
although the exact shape of the IMF has been characterized by similar power-law slopes, 
log-normal distributions, or Gaussian distributions (e.g., Miller \& Scalo 1979; 
Gilmore, Reid, \& Hewett 1985; Scalo 1986; Hawkins \& Bessell 1988; Stobie, Ishada, \& Peacock 
1989; Scalo 1998; Kroupa, Tout, \& Gilmore 1993; Kroupa 2001; 2002; Reid, Gizis, \& Hawley 2002).  
At a characteristic mass that is $<$1~$M_\odot$, several of these studies have reached the conclusion 
that the IMF flattens.  For example, Kroupa et~al.\ measure a break in the IMF slope at 0.5~$M_\odot$ from 
$\alpha$ = $-$2.3 to $\alpha$ = $-$1.3.

IMF studies that are based on the local luminosity function offer both 
advantages and disadvantages over alternative methods.  The most robust data come from 
stars within a few tens of parsecs, where distances are well measured from parallaxes and 
binarity is resolved.  However, limited sample sizes lead to strong statistical errors 
over specific regions of the IMF we seek to constrain.  Expanded samples of disk 
stars are available from wide-field photometric surveys, but the resulting luminosity 
functions suffer from strong Malmquist bias and disagree with the nearby sample unless 
detailed corrections are made (see e.g., Kroupa 1995 and Reid \& Gizis 1997 for 
discussions).  Other methods to measure the IMF include modeling the luminosity functions 
of star forming regions, young star clusters, and old clusters.   The nature of these stellar 
systems as simple populations offers a tremendous advantage over field studies.  
Over a wide spectrum of mass, the constituent stars share incredible similarities in 
their (well measured) properties.  Unfortunately, derivation of the IMF from such populations 
is affected by a different set of errors that are often difficult to assess (e.g., photometric 
uncertainties due to high levels of extinction in star forming regions, membership errors due to 
field interlopers in more sparse populations, and mass-segregation effects in dynamically-relaxed 
older clusters).  Further discussion of recent measurements of the IMF from clusters is 
provided in the review by Bastian, Covey, \& Meyer (2010).  

We present a new derivation of the stellar IMF based on high-precision {\it Hubble Space 
Telescope (HST)} photometry and astrometry of the outskirts of the Small Magellanic Cloud (SMC).  
The population represents an independent tool to bear on the study of the IMF, and 
offers several advantages over previous studies.   First, photometry above the 75\% completeness 
limit extends from $F606W$ = 22.6 to 28.7 and includes $>$5,000 stars on the unevolved main 
sequence, thereby providing a high-resolution mapping of the complete stellar mass distribution 
between $M$ = 0.37 -- 0.93~$M_\odot$.  Second, the population forms a tight sequence on the 
color-magnitude diagram (CMD) similar to a star cluster, and is therefore approximately co-spatial 
and also only contains a small metallicity spread.  Population members are selected from 
high-precision proper motions.  Finally, the field SMC stars are well mixed dynamically, so the 
present day mass function can be assumed to be consistent with the IMF.\footnote{We refer to the 
{\it initial} mass function as the distribution of stellar masses following the process of 
star formation.  This is not to be confused with the ``primordial'' mass function at 
high-redshift.}


\section{Observations and Data Reduction} \label{sec:data}

The primary science goal motivating these observations was to characterize the complete stellar populations 
of the nearby Milky Way globular cluster 47~Tuc, in order to derive the cluster white dwarf cooling 
age (Hansen et~al.\ 2013, in prep.).  However, the line of sight through 47~Tuc also intersects the 
outskirts of the Small Magellanic Cloud (SMC) dwarf galaxy.  At the location of the specific 47~Tuc 
field, $\alpha$ = 00:22:39 and $\delta$ = $-$72:04:04, the stellar populations surveyed in the SMC 
are $\sim$2.3~degrees (2.4~kpc) west of the galaxy center.

The observations were obtained over 121 orbits of exposure time with {\it HST} and the 
Advanced Camera for Surveys (ACS) in GO-11677 (PI: H.\ Richer).  We obtained 117 exposures in $F606W$ 
(163.7~ks) and 125 exposures in $F814W$ (172.8~ks).  These two filters provide superb sensitivity at optical 
wavelengths, and are ideally suited for high signal-to-noise ratio photometry of the SMC main sequence.  The 
imaging field was observed at 13 different orientations, each separated by $\sim$20~degrees, and is 
therefore well-dithered to enable resampling of the point spread function (PSF).  The effective angular size 
of the observations is 5.25$'$ $\times$ 5.25$'$.

The data reduction for these observations is described in detail in Kalirai et~al.\ (2012).  Briefly 
summarizing, we first corrected all of the images for charge transfer inefficiency using the pixel-based 
corrections from \cite{anderson10}.  We then generated distortion-free images using MultiDrizzle \citep{fruchter97}, 
and calculated transformations between each of these images to link them to a reference frame in each 
filter.  The transformations were based on Gaussian-fitted centroids of hundreds of stars on each image (i.e., 
47~Tuc and SMC stars), and the solution was refined through successive matches.  The final offsets provide 
alignment of the individual images to better than 0.01 pixel.  A second pass of MultiDrizzle was performed 
on the aligned images to flag cosmic rays and hot pixels.  For this step, the sky background was calculated 
for each individual image and offsets were made to normalize the values.  A third pass of MultiDrizzle was 
executed on the clean images to create supersampled stacks with a pixel scale of 0.03~arcsec, giving a PSF 
with a full-width half-maximum (FWHM) of 2.7~pixels.


\begin{figure}[t]
\begin{center}
\leavevmode 
\includegraphics[width=8.5cm]{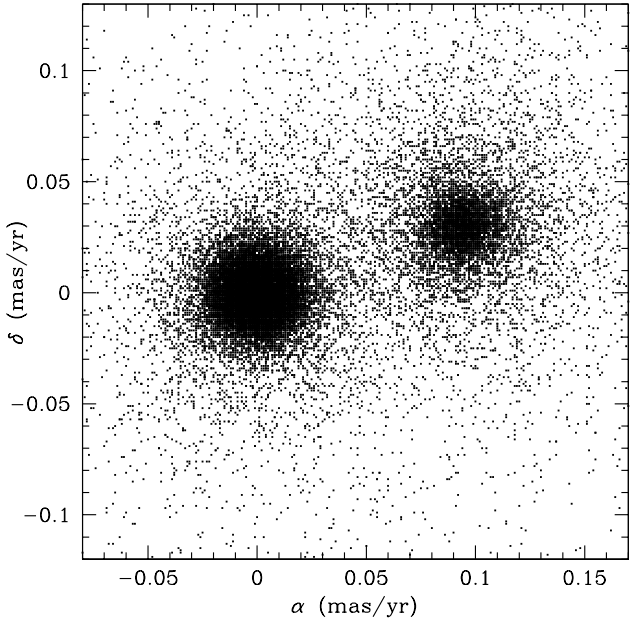}
\end{center}
\vspace{-0.5cm}
\caption{The proper motion distribution of stars along the sightline, based on a $\sim$6-year baseline 
between the present observations and a large number of {\it HST} archival images.  The SMC population 
(right clump) is clearly separated from the cluster population. \label{fig:PMs}}
\end{figure}


To measure the photometry and morphology of all sources, the stand alone versions of the 
DAOPHOT~II and ALLSTAR photometry programs were used on the stacked images \citep{stetson87,stetson94}.  The 
final catalog is based on first performing aperture photometry on all sources that are at least 
2.5$\sigma$ above the local sky, then deriving a PSF from $\sim$1000 high signal-to-noise ratio and isolated 
stars in the field, and finally applying the PSF to all sources detected in the aperture photometry list.  
The PSF was calculated using a multi-step iterative method that built up to allow for third order polynomial 
spatial variations across the field.  The final catalog contains sources that were iteratively matched between 
the two images, and cleaned to eliminate background galaxies with $\chi^2$ and sharpness cuts from the PSF 
fitting.

The location of the specific field that we targeted in 47~Tuc was chosen to overlap a large number 
of archival {\it HST} calibration images.  The average date of these data is 2004, so this presents a 
6-year baseline over which proper motions can be measured.  Positions of all stars in these archival 
data were derived, and cross-identified with the new 2010 observations.  Over the baseline, the SMC 
population is clearly separated from the 47 Tuc population, as shown in Figure~\ref{fig:PMs}.  
Detailed information on the proper motion measurements, including studies of the bulk motions of 
47~Tuc and the SMC, as well internal motions of individual stars within these populations, will 
be provided in future papers.  Here, we use the proper motions to cleanly separate the 
SMC stellar population from both the foreground 47~Tuc stars and from any residual extragalactic 
contamination caused by faint (nearly unresolved) galaxies.


\begin{figure*}[ht]
\begin{center}
\leavevmode 
\includegraphics[width=12cm, angle=270, bb = 22 16 596 784]{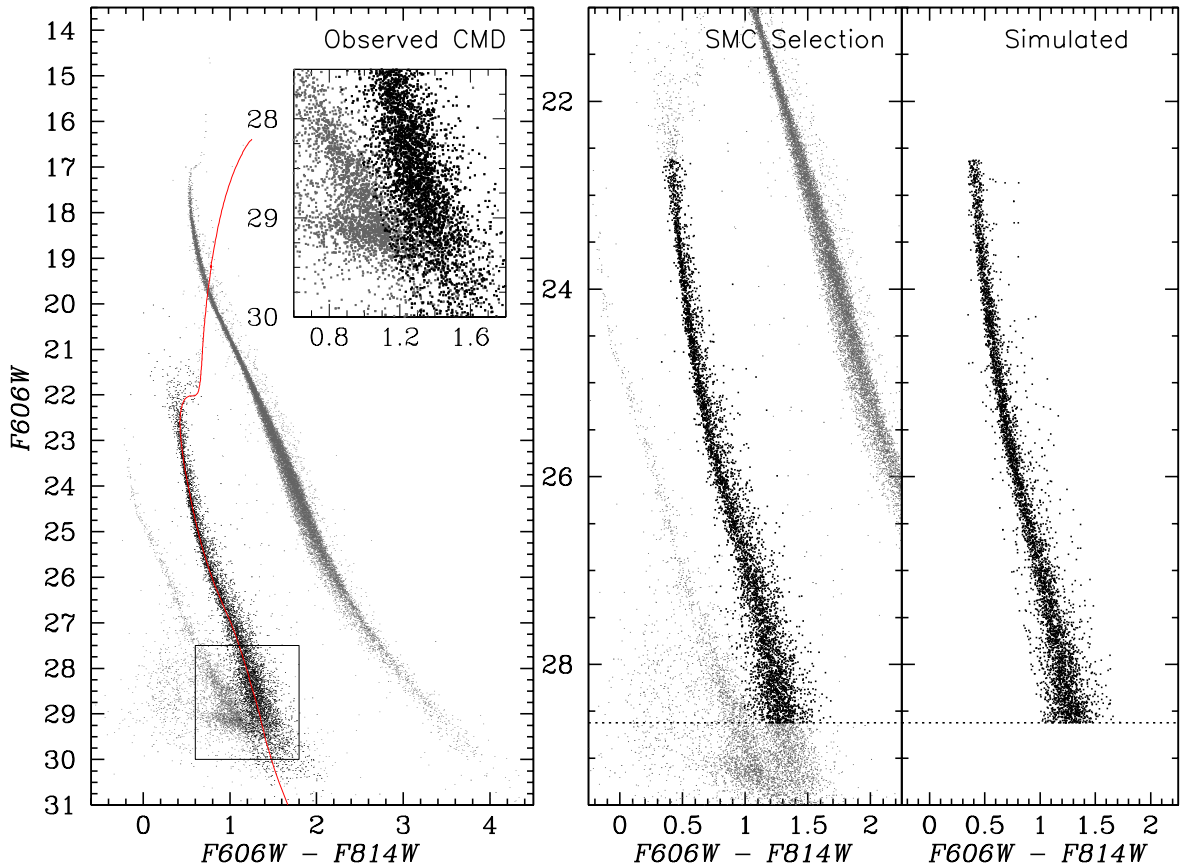}
\end{center}
\vspace{-0.5cm}
\caption{The CMD of all stars along the line of sight in the {\it HST}/ACS observations (left).  The 
photometry reveals three distinct populations; the main sequence and white dwarf cooling sequence of the foreground 
Milky Way globular cluster 47~Tuc (grey points), and the main sequence of the outskirts of the background SMC dwarf 
galaxy (i.e., the middle sequence -- darker points).  With a 50\% completeness limit of $F606W$ = 29.9 
(see \S\,\ref{sec:art}), this imaging represents the deepest probe of the SMC's populations to date.  The inset panel 
presents a closer view of the faintest stars on the lower SMC main sequence (dark points) and 47 Tuc white dwarf cooling 
sequence (grey points).  The red curve illustrates a stellar isochrone with [Fe/H] = $-$1.1 and $t$ = 7~Gyr 
\citep{dotter08}, at a distance of ($m - M$)$_0$ = 19.1 $\pm$ 0.1 (determined directly, as described in 
\S\,\ref{sec:agemetallicity}).  The proper-motion selected SMC population, down to the limit of the data, is shown 
as darker points in the left panel, and reproduced down to $F606W$ = 28.6 in the middle panel.  As discussed in 
\S\,\ref{sec:IMF}, we simulate the SMC population by drawing stars from an IMF and convolving them with 1.) a range 
of mass-luminosity relations appropriate for the SMC halo, 2.) binaries, 3.) photometric scatter in the observations, 
and 4.) incompleteness in the observations.  An example of the simulated CMD, for an IMF with form $dN/dM \propto$ $M^{-1.90}$, 
is shown in the right panel. \label{fig:cmd3panel}}
\end{figure*}



\section{Analysis} \label{sec:analysis}

We derive the IMF through several steps.  First, we measure the luminosity function of the SMC by isolating 
the stellar main-sequence of the galaxy using the proper motions.  Next, we generate simulations of this population.  
This is done by drawing masses from different IMFs, and interpolating the masses within a small grid of stellar 
isochrones (i.e., mass-luminosity relations) to yield magnitudes.  These ages and metallicities of these isochrones 
are chosen to match the CMD of the SMC population.  This analysis includes careful consideration to ensure 
the simulated SMC population accounts for the photometric scatter and incompleteness of the real data.  At 
this point, luminosity functions are constructed for the simulated main sequence and compared to the 
observations.

\subsection{Isolating the SMC Stellar Population} \label{sec:isolate}

Photometry of all stars along the sightline is illustrated in Figure~\ref{fig:cmd3panel}.  The CMD in the left 
panel reveals three populations of stars; the 47~Tuc main sequence (reddest stars), the SMC main sequence (middle), 
and the 47~Tuc white dwarf cooling sequence (bluest stars).  In color-magnitude space, there is a mild 
separation between the latter two populations down to the faint limit of the data.  This is illustrated in more 
detail in the small inset panel that focuses on the faint part of the CMD near $F606W$ = 29.25.  The darker points 
are selected as SMC members based on their proper motion (see Figure~\ref{fig:PMs}), and are reproduced in the middle 
panel on a finer scale down to $F606W$ = 28.6 (this limit is discussed in \S\,\ref{sec:IMF}).   

The luminosity function of the SMC based on the selection described above is provided in Table~1.  Both 
the raw counts from the confirmed members of the SMC and the completeness corrected star counts, 
based on the artificial star tests described below, are presented.  

\subsection{Metallicity, Distance, and Age} \label{sec:agemetallicity}

The complete SMC main sequence in this halo field is measured down to $F606W \gtrsim$ 30.  This is, therefore, the 
deepest investigation of the CMD of the galaxy to date.  Previous wide-field 
photometric analysis of the field populations of the SMC suggest that the galaxy formed half of its stars $>$8~Gyr ago 
(e.g., Harris \& Zaritsky 2004), and experienced more recent epochs of star formation in the main body.  Surveys of 
the ``halo'' of the galaxy have characterized the bulk of the population as being old, although there are small 
differences in the measured star formation history among different regions (e.g., Graham 1975 -- a field near 47~Tuc; 
Dolphin et~al.\ 2001; Noel et~al.\ 2007; Sabbi et~al.\ 2009).  Both photometric and spectroscopic studies across the halo 
of the SMC indicate evidence for a radial metallicity gradient (e.g., Sabbi et~al.\ 2009; Carrera et~al.\ 2008).  In the 
western direction toward 47~Tuc at a distance $>$1.5~kpc, the metallicity of the population is [Fe/H] $\lesssim$ $-$1.1 
(Carrera et~al.\ 2008).  

To model the SMC population as a distribution of masses, we first compare the stellar main sequence on the CMD
to a grid of moderately metal-poor stellar isochrones from the updated models of Dotter et~al.\ (2008).  
Although the distance to the main body of the SMC is now precisely measured to be 60.6 $\pm$ 1.0 $\pm$ 
2.8~kpc based on the analysis of 40 eclipsing binaries (Hilditch, Howarth, \& Harries 2005), there exists a 
significant depth to the galaxy of up to 20~kpc \citep{mathewson88,hatzidimitriou93,crowl01,lah05,haschke12}.  
Specifically, the eastern regions are closer than the western regions, so we expect the SMC population along 
our line of sight to have a larger distance than the main body.  Fortunately, we can measure the distance to the 
population directly by using the unevolved fiducial of the bright 47~Tuc main sequence, a metal-poor population at 
($m - M$)$_0$ = 13.36 $\pm$ 0.06 (see Woodley et~al.\ 2012, and references therein).  By aligning the 
two sequences at $F606W$ = 24 -- 27, we derive the distance modulus of the SMC halo along this line of 
sight to be ($m - M$)$_0$ = 19.1 $\pm$ 0.1.  This calculation includes an offset of $\Delta$($F606W$) = 
$-$0.20 -- 0.30 to translate the [Fe/H] = $-$0.8 star cluster population to a range of SMC metallicities with 
$-$1.4 $<$ [Fe/H] $<$ $-$1.0 (derived from the Dotter et~al. models -- see below).  The uncertainty 
in the distance modulus is taken to be the small range over which we achieve acceptable fits of the 
two sequences, gauged by eye.  The derived distance to the SMC population along this line of sight is 
in good agreement with the recent three-dimensional maps of the SMC based on RR Lyrae stars (e.g., 
top-left panel of Figure~5 in Haschke, Grebel, \& Duffau 2012).

In Figure~\ref{fig:cmd3panel}, we illustrate a stellar isochrone with [Fe/H] = $-$1.1 and 
$t$ = 7~Gyr superimposed on the observed SMC population.  The isochrone is our best fit to the data, 
and nicely reproduces the main sequence, turnoff, sub-giant branch, and red-giant branch.  According to this 
mass-luminosity relation, the faintest detected SMC stars at $F606W$ = 30.5 have $M$ = 0.17~$M_\odot$.  The 
``thickness'' of the SMC main sequence indicates that the stellar population is both co-spatial and contains only 
a small metallicity spread.  For example, a direct comparison of the SMC sequence to that of 47~Tuc (a simple stellar 
population) on the same CMD indicates almost no additional broadening other than what is expected from photometric 
scatter (and binaries).  Formally, we can rule out models with metallicities that fall outside of 
$-$1.4 $<$ [Fe/H] $<$ $-$1.0 given the combined high-precision {\it HST} observations of the SMC lower 
main sequence and red-giant branch.  To account for this small metallicity spread in our derivation of 
the stellar IMF, our analysis below uses stellar isochrones (i.e., mass-luminosity relations) populated 
within this range.  


\begin{figure*}[ht]
\begin{center}
\leavevmode 
\includegraphics[width=6.0cm, angle=270, bb = 22 16 330 784]{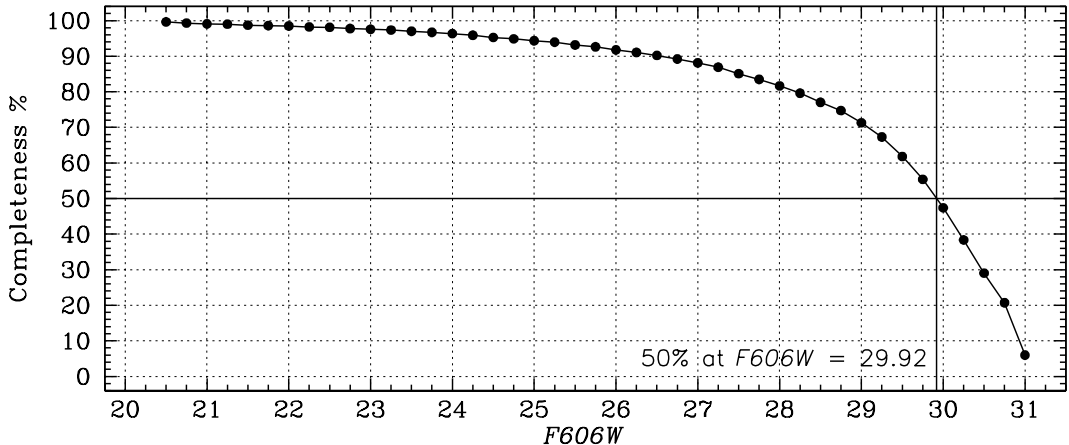}
\end{center}
\vspace{-0.2cm}
\caption{The completeness fraction of the artificial star test experiments on the {\it HST} 
images of the SMC population.  As discussed in \S\,\ref{sec:art}, input stars were placed on 
the SMC main sequence in the CMD and recovered blindly using the same photometric algorithms 
that were applied to the real data.  A total 8000 trials were generated, with 4~million 
artificial stars (i.e., about $\sim$100,000 in each 0.25 mag bin).  The 50\% completeness 
limit of this data set, measured by requiring the star to be found in both filters, is 
$F606W$ = 29.9. \label{fig:art}}
\end{figure*}


Before continuing, we note that the age and metallicity of this remote population in a disturbed galaxy is 
interesting to study in its own right, for example, to establish the level of interaction-driven star 
formation and to constrain halo assembly models.  The morphology of the main-sequence turnoff in the cleaned 
photometry of Figure~\ref{fig:cmd3panel} (left) shows multiple splittings, and so even this remote field of 
the SMC halo is not co-eval.  The brighter main-sequence turnoffs and sub-giant branches extend to $F606W$ 
$\sim$ 21.2, with the most dominant population having $t$ = 4.5~Gyr based on the same grid of stellar 
models.  To avoid any biases from missing massive stars that have evolved in the younger populations, our 
analysis will only include stars on the SMC main sequence that are fainter than the oldest turn 
off.  At such old ages, the relation between mass and luminosity for these unevolved stars has a negligible 
dependence on age.

\vspace{1.0cm}

\subsection{Correcting Data Incompleteness} \label{sec:art}

To measure the IMF from the distribution of stars along the SMC main sequence, we model 
the observations as a convolution of the input stellar masses, the mass-luminosity relation, 
and the selection functions in the data.  We characterized the latter by generating an extensive 
set of artificial star tests and analyzing these to constrain the data incompleteness and 
photometric errors.  First, the stellar PSF was used to 
generate stars over the complete luminosity range occupied by real stars on the SMC main sequence 
(a flat luminosity function was used).  These stars were injected into each of the $F606W$ and 
$F814W$ images simultaneously, with a color consistent with the observed SMC main sequence.  The 
fraction of stars injected into each image was set to $\sim$1\% of the total number of stars in 
the image, so as to not introduce incompleteness due to crowding in the tests themselves.  A 
total of 8000 trials were generated, producing 4~million artificial stars.


\begin{table}
\begin{center}
\caption{The SMC Luminosity Function}
\begin{tabular}{lcc}
\hline
\hline
\multicolumn{1}{c}{F606W} & \multicolumn{1}{c}{No. Stars} & \multicolumn{1}{c}{No. Stars} \\
\multicolumn{1}{c}{} & \multicolumn{1}{c}{(raw)} & \multicolumn{1}{c}{(corr)} \\ 
\hline
22.75 ($\pm$ 0.125) &  89  $\pm$  9 &   91.0 $\pm$  9.7  \\
23.00 ($\pm$ 0.125) &  90  $\pm$  9 &   92.2 $\pm$  9.7  \\
23.25 ($\pm$ 0.125) &  93  $\pm$ 10 &   95.5 $\pm$  9.9  \\
23.50 ($\pm$ 0.125) & 115  $\pm$ 11 &  118.5 $\pm$ 11.1  \\
23.75 ($\pm$ 0.125) & 147  $\pm$ 12 &  152.0 $\pm$ 12.5  \\
24.00 ($\pm$ 0.125) & 137  $\pm$ 12 &  142.2 $\pm$ 12.1  \\
24.25 ($\pm$ 0.125) & 154  $\pm$ 12 &  160.6 $\pm$ 12.9  \\
24.50 ($\pm$ 0.125) & 169  $\pm$ 13 &  177.4 $\pm$ 13.6  \\
24.75 ($\pm$ 0.125) & 166  $\pm$ 13 &  175.0 $\pm$ 13.6  \\
25.00 ($\pm$ 0.125) & 175  $\pm$ 13 &  185.5 $\pm$ 14.0  \\
25.25 ($\pm$ 0.125) & 184  $\pm$ 14 &  195.8 $\pm$ 14.4  \\
25.50 ($\pm$ 0.125) & 167  $\pm$ 13 &  179.2 $\pm$ 13.9  \\
25.75 ($\pm$ 0.125) & 189  $\pm$ 14 &  204.0 $\pm$ 14.8  \\
26.00 ($\pm$ 0.125) & 204  $\pm$ 14 &  222.3 $\pm$ 15.6  \\
26.25 ($\pm$ 0.125) & 190  $\pm$ 14 &  208.7 $\pm$ 15.1  \\
26.50 ($\pm$ 0.125) & 215  $\pm$ 15 &  238.3 $\pm$ 16.3  \\
26.75 ($\pm$ 0.125) & 207  $\pm$ 14 &  232.0 $\pm$ 16.1  \\
27.00 ($\pm$ 0.125) & 226  $\pm$ 15 &  256.4 $\pm$ 17.1  \\
27.25 ($\pm$ 0.125) & 272  $\pm$ 16 &  313.0 $\pm$ 19.0  \\
27.50 ($\pm$ 0.125) & 265  $\pm$ 16 &  311.4 $\pm$ 19.1  \\
27.75 ($\pm$ 0.125) & 294  $\pm$ 17 &  352.2 $\pm$ 20.5  \\
28.00 ($\pm$ 0.125) & 354  $\pm$ 19 &  433.5 $\pm$ 23.0  \\
28.25 ($\pm$ 0.125) & 422  $\pm$ 21 &  530.0 $\pm$ 25.8  \\
28.50 ($\pm$ 0.125) & 435  $\pm$ 21 &  564.6 $\pm$ 27.1  \\
\hline
\normalsize
\end{tabular}
\end{center}
\end{table}

Each of these new images, one per trial in each filter, were subjected to the same photometric 
routines that were applied to the actual drizzled images, using identical criteria.  The stars 
were recovered blindly and automatically cross-matched between filters and to the input 
star lists containing actual positions and fluxes.  Stars that were not recovered were also retained 
in the final matched lists and flagged as such.  The end result of this process is a large scattering 
matrix that defines the fidelity of the observations and data reduction, including the photometric 
error distribution at any point in the CMD and the completeness.  The 50\% completeness limit 
of the SMC population in the joint $F606W$, $F814W$ CMD is $F606W$ = 29.9 (see Figure~\ref{fig:art}).  

We also assess whether the proper motion selection imparts any additional incompleteness in 
the SMC population.  An independent way to test this is to consider the stellar main sequence of 
47~Tuc, which extends to beyond 30th magnitude in a region of color-magnitude space where there are 
no background galaxies.  By comparing the luminosity functions with and without proper motion selection, 
we confirm that there is no additional incompleteness down to at least $F606W$ = 29, and possibly 
some very minor incompleteness fainter than this (e.g., much smaller than the data incompleteness 
itself).

\subsection{The Initial Mass Function of Stars} \label{sec:IMF}

We simulate the CMD in Figure~\ref{fig:cmd3panel} to derive the IMF through a multi-step process.  First, random 
masses are drawn from a power-law IMF with a specific slope and populated within each of the five 
$-$1.4 $<$ [Fe/H] $<$ $-$1.0 stellar isochrones (i.e., see Figure~\ref{fig:cmd3panel} for the [Fe/H] = $-$1.1 
model).  For each bandpass, this provides luminosities for each mass at each metallicity.  The results from 
the five metallicities are combined together with equal weighting.  The simulation accounts for binaries by 
drawing a secondary companion star from the 
IMF for a fraction of the population, and setting the luminosity of the unresolved binary as the sum of the 
light from the two stars.  The binary fraction among low mass stars such as those in our study is 25 -- 35\% 
(Leinert et~al.\ 1997; Reid \& Gizis 1997; Delfosse et~al.\ 2004; Burgasser et~al.\ 2007), and so we set the 
fraction to 30\% in our simulation (see below).  Next, for each mass in the input IMF, we 
randomly select stars from the $F606W$, $F814W$ artificial star cloud distribution if the luminosities associated 
with the input mass are consistent with the {\it input} magnitudes of a star in the cloud to 0.05 magnitudes 
(in both filters).  The luminosities that are extracted from the cloud are the {\it output} magnitudes of the matches.  
This method accounts for the photometric scatter and completeness of the data.  The resulting simulated SMC CMD is 
shown in Figure~\ref{fig:cmd3panel} (right).


\begin{figure}[t]
\begin{center}
\leavevmode 
\includegraphics[width=8.5cm]{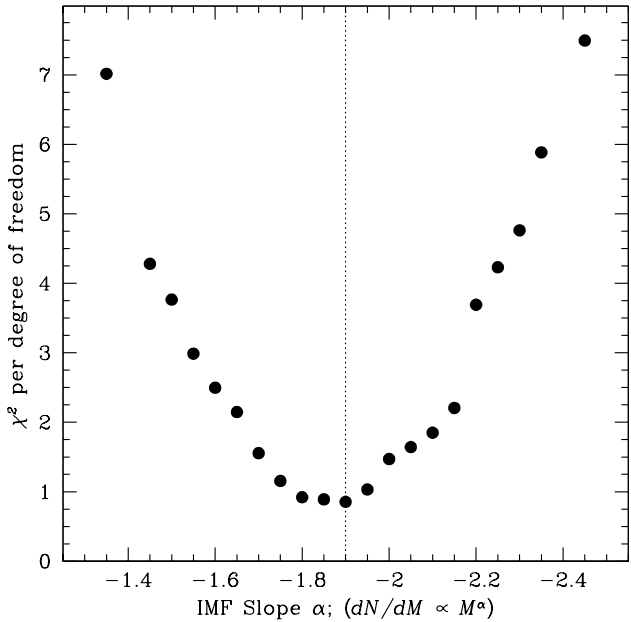}
\end{center}
\vspace{-0.5cm}
\caption{The distribution of $\chi$$^2$ from the fit of power-law IMFs to the observed SMC luminosity 
function.  The minimum occurs at $\alpha$ = $-$1.90 ($^{+0.15}_{-0.10}$) (3$\sigma$ error), and 
the $\chi$$^2$ per degree of freedom for this fit is 0.86.\label{fig:ChiSqPLOT}}
\end{figure}



\begin{figure*}[ht]
\begin{center}
\leavevmode 
\includegraphics[width=9.0cm, angle=270, bb = 22 16 430 784]{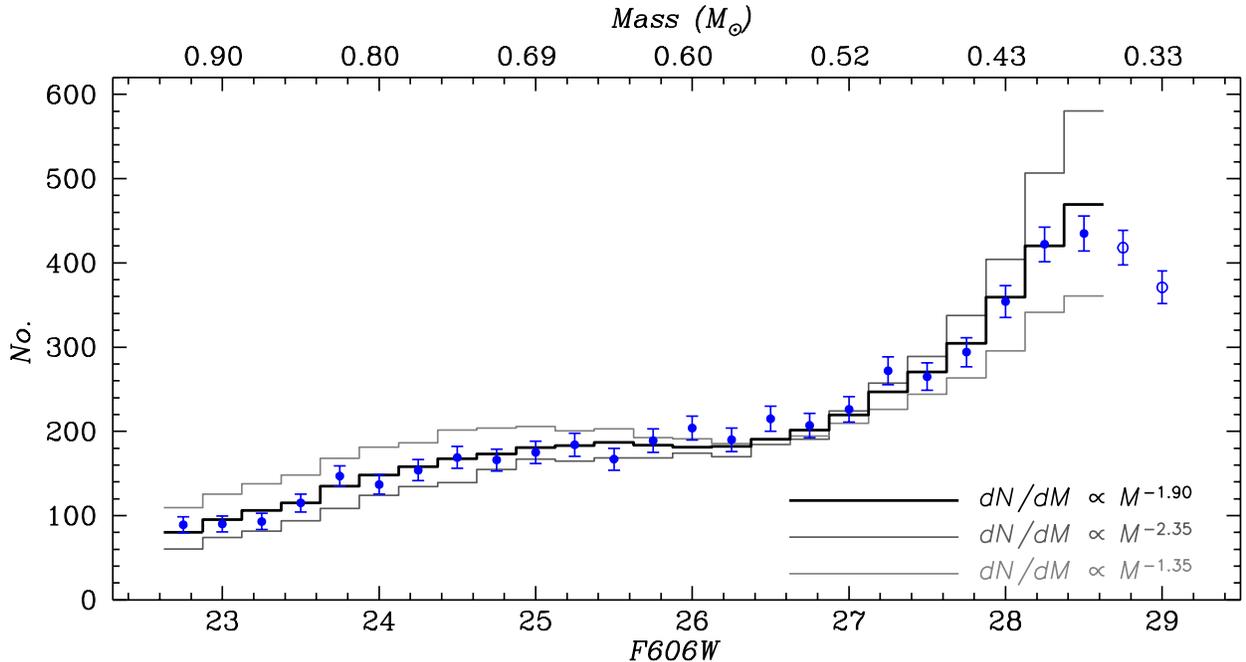}
\end{center}
\vspace{-0.5cm}
\caption{The deep luminosity function of the SMC from our {\it HST}/ACS observations is illustrated 
with blue points and error bars, along with three of the simulated luminosity functions based on 
different input IMFs (black and grey curves).  The simulations are produced as described in 
\S\,\ref{sec:IMF}.  We obtain an excellent fit to the observations over the entire luminosity 
range extending from below the main-sequence turnoff to $F606W$ = 28.6 with a single power law.  The 
best-fit IMF slope over this range, from $M$ = 0.37 -- 0.93~$M_\odot$, is $\alpha$ = $-$1.90 (black curve).  
For comparison, the two grey curves illustrate IMFs with $\alpha$ = $-$1.35 (lower counts at low masses) 
and $-$2.35 (the Salpeter IMF; higher counts at low masses).  Fainter than $F606W$ = 28.6, the observed SMC 
luminosity function shows a turnover which would not be reproduced by an extension of the $\alpha$ = $-$1.90 
power law.
\label{fig:MF}}
\end{figure*}


To measure the IMF, we compute the reduced chi square statistic ($\chi$$^2$) between the luminosity 
function of each of the simulations and the SMC observations.  The distribution of reduced $\chi$$^2$ is 
shown in Figure~\ref{fig:ChiSqPLOT}, and exhibits a minimum at $\alpha$ = $-$1.90 ($^{+0.15}_{-0.10}$) 
(3$\sigma$ error). The $\chi$$^2$ per degree of freedom is 0.86 for this IMF (i.e., $\chi$$^2$ = 
19.7 for 24 degrees of freedom).  The best-fit power-law 
form of the stellar IMF from these data is therefore $dN/dM \propto$ $M^{-1.90}$, shallower than the 
Salpeter slope of $\alpha$ = $-$2.35.  A comparison of this derived IMF to the observed SMC luminosity function 
is shown in Figure~\ref{fig:MF}, and demonstrates excellent agreement over the entire range of masses from 
$M$ = 0.37 -- 0.93~$M_\odot$ (black curve).  Figure~\ref{fig:MF} also illustrates two power-law IMFs with 
slopes that bracket the best-fit value as grey curves.  This includes a shallower slope of $\alpha$ = 
$-$1.35 (smaller counts at faint magnitudes) and the Salpeter slope of $\alpha$ = $-$2.35, neither of which 
agree with the data.  

In this analysis, the observed luminosity function of the SMC is fit down to $F606W$ = 28.6 (i.e., the 
limit shown in Figure~\ref{fig:cmd3panel}).  Below this limit, the data indicate some evidence for a 
turnover in the luminosity function that is inconsistent with an extension of this power law (i.e., 
the open circle points in the last two bins).  It is difficult to constrain the slope of the 
mass function that matches these data given the limited leverage in stellar mass.  Therefore, we stress 
that our primary result of an IMF with form $dN/dM \propto$ $M^{-1.90}$ is only valid down to $M$ = 0.37~$M_\odot$, 
and should not be extrapolated to lower masses.


\section{Discussion} \label{sec:discussion}

The derivation of the stellar IMF at $M <$ 1~$M_\odot$ gives a best-fit slope that is slightly shallower than a Salpeter 
IMF, which itself has had success in reproducing the observed luminosity function of nearby populations (the Salpeter 
1955 result of $\alpha$ = $-$2.35 is valid down to $M$ = 0.4~$M_\odot$).  More importantly, the nature of our 
study provides excellent leverage to constrain the shape of the IMF over a {\it continuous} mass range with 
$M \lesssim$ 1~$M_\odot$, and is based on $>$5,000 stars.  Several previous studies have suggested that the 
IMF exhibits a ``break'' within this mass range.  For example, the \cite{chabrier03} log-normal IMF exhibits a 
shallow turnover in this mass range, the \cite{kroupa01} power-law IMF indicates a slope change from $-$2.3 to $-$1.3 at 
$M$ = 0.5~$M_\odot$, and the \cite{reid02} analysis suggests that a break from a steeper IMF occurs at 
$M$ = 0.7 -- 1.1~$M_\odot$.  More recently, \cite{bochanski10} measured the IMF using multi-band photometry 
from the Sloan Digital Sky Survey (SDSS) over 8400 deg$^2$ ($\sim$15 million stars), and found that the stellar 
distribution between $M$ = 0.32 -- 0.8~$M_\odot$ is consistent with a single power-law ($\alpha$ = $-$2.4) and is 
significantly shallower at lower masses (i.e., or a log-normal distribution with $M_0$ = 0.25 $M_\odot$).

The analysis of SMC stars with $M$ = 0.37 -- 0.93~$M_\odot$ suggests that we do not require a two-component 
IMF to represent the data.  A single power law is the simplest form of the stellar IMF, and it reproduces 
the observations nicely.  Formally, if we fit a subset of the SMC population that only includes stars with 
$M >$ 0.60~$M_\odot$ (i.e., a cut at $F606W \lesssim$ 26, the mid point of the luminosity range), the best-fit 
slope is $\alpha$ = $-$2.05 $\pm$ 0.3 (with a much flatter $\chi$$^2$ distribution).  This is therefore in 
excellent agreement with our single power law over the full mass range.  If we force a broken power-law fit, 
acceptable matches to the data are only obtained if the transition mass is shifted to fairly low masses 
$M <$ 0.5~$M_\odot$ {\it and} the difference in slope between the top and bottom ends is small (i.e., 
within $\pm$0.3 of $\alpha$ = $-$1.90).  

Many of the previous studies of the IMF that are based on local Milky Way field stars have relied on subsets of 
the same sample of stars.  The present study is independent of these previous analysis, and of a very different 
nature.  Our method takes advantage of a sample of SMC stars that are co-spatial and that share similarities, yet it 
requires corrections to deal with binaries, photometric uncertainty, and completeness.  Modest changes in the 
binary fraction do not affect our results.  For example, we computed a separate set of IMFs with binary 
fractions of 20\% and 40\%, and found the best-fit IMF to the observed SMC luminosity function to be within $\pm$0.05 
of the $\alpha$ = $-$1.90 slope (the slope is steeper if the binary fraction is $\lesssim$10\%).  The completeness and 
photometric scatter is directly derived from an extensive 
set of artificial star tests as described in \S\,\ref{sec:art}, and folded into our simulations to ensure no biases 
are introduced.  Specifically, the SMC-selection from the CMD is very robust over the magnitude range adopted, and the 
completeness corrections are not large.  Other systematics related to errors in the derived distance of the population 
or an error in the mass-luminosity relation of stars could have a minor impact on the slope of the derived IMF, but is 
unlikely to lead to a (large) systematic flattening of the slope over a specific mass range.  

\section{Implications} \label{sec:implications}

\subsection{Theory of Star Formation}

There are several implications of our result.  First, studies of the IMF can constrain aspects of star formation 
theory.  In one popular model, the mass distribution that is predicted from star formation is 
related to the efficiency of fragmentation in molecular clouds.  Numerical simulations of turbulence and the 
resulting shock velocities (e.g., Larson 1979; 1981) indicate that the clump mass distribution is similar 
to a Salpeter-type (steep) IMF at $M$ $\gtrsim$1~$M_\odot$ (Padoan \& Nordlund 2002).  Although the masses of 
the clumps continue to follow a power-law distribution down to lower masses (i.e., for scale-free turbulence), 
the IMF results from only those cores that are dense enough to collapse.  The measurement of a slope for the IMF 
that is shallower than the Salpeter slope for $M$ = 0.37 -- 0.93~$M_\odot$ (and yet shallower below this mass limit) 
suggests that a lower fraction of the low mass cores suffered gravitational collapse and formed stars.  This scenario 
suggests that the gas density in the molecular clouds is lower than what would be needed if the IMF continued at the 
Salpeter slope down to low masses (e.g., see models in Figure~1 of Padoan \& Nordlund 2002).

Other star formation models involving gravitational fragmentation followed by a balance between accretion and 
dynamical ejection also predict specific shapes for the IMF.  For example, \cite{bate05} use hydrodynamical 
calculations to demonstrate that the characteristic mass of the IMF is dependent on the mean thermal Jeans mass 
of the clouds, $M_{\it Jeans}$.  Our finding of a shallower mass function at $M$ $<$ 1~$M_\odot$ would be 
indicative of a higher $M_{\it Jeans}$, and therefore less dense clouds relative to predictions from steeper 
IMFs (see Moraux et~al.\ 2007 and references therein for further discussion).  The exact impact of this finding 
depends critically on the shape of the IMF at masses that are lower than we measure here.

\subsection{The Mass Budget of the Milky Way and $M/L$ Ratios}

Two of the most widely used applications of the IMF are to constrain the mass budget of the Milky Way and the 
mass-to-light ratios ($M/L$) of unresolved galaxies.  Whereas higher mass stellar sources are responsible for 
the bulk of the energetics in galaxies and the light output, it is precisely the shape of the IMF at 
$M$ $<$ 1~$M_\odot$ that dominates the inferred galaxy masses.  Previous analysis has frequently adopted a 
Salpeter IMF over a broad mass range from 0.1 -- 100~$M_\odot$ (i.e., outside the limits that it was 
constrained).  For an old, simple stellar population, this leads to $M/L$ ratios that are about a factor of 
two larger than the Chabrier IMF (i.e., see Chabrier 2003).  For this generic example, our IMF would 
provide a $M/L$ ratio that is lower than these IMFs given the shallower slope at intermediate masses, 
with the exact value being dependent on the assumption of the IMF at $M$ $<$ 0.37~$M_\odot$.  The full impact 
of the new IMF on understanding the properties of unresolved light from distant galaxies will require 
in-depth calculations in population synthesis techniques (e.g., Bruzual \& Charlot 2003).  As a related 
example, future estimates of the total mass of resolved populations from CMD analysis will be affected by 
our findings.  As the lower mass stellar population is often undetected, an extrapolation with a single 
(shallower) power-law form down to $M$ = 0.37~$M_\odot$ will yield different stellar masses than a steeper 
IMF with a break at higher masses.

\subsection{Metallicity Variations}

Understanding any variation of the IMF with metallicity is fundamental, since $M_{\it Jeans}$ depends 
on temperature and density, which in turn depend on processes that are linked to metallicity (e.g., cooling 
and dust emission).  Despite many contrary remarks in the literature, the variation of the IMF with metallicity 
is poorly constrained.  For example, the comparison of stellar luminosity functions in various star clusters with 
different metallicities requires specific corrections due to incompleteness in the data set and for the 
dynamical state of the particular cluster.  Different investigators also use preferred sets of models 
and modeling techniques, thereby making it difficult to assess the systematic errors in the measurements.
Recent work by van Dokkum \& Conroy (2012) and Conroy \& van Dokkum (2012) suggest that a true variation 
in the IMF does exist. Their work takes advantage of gravity sensitive absorption lines in the integrated 
light of old stellar populations of early type galaxies, where they find that galaxies with deeper potential 
wells have more dwarf-enriched mass functions (i.e., a more bottom heavy mass function).  \cite{conroy12} 
suggest that the IMF is steeper than a Salpeter IMF in early-type galaxies with the highest dispersions and 
[Mg/Fe].  

The SMC stellar population offers a new opportunity to establish high-precision measurements of the IMF over 
an appreciable mass range with $M$ $<$ 1~$M_\odot$, at a well determined metallicity.  Our constraints on 
the stellar IMF from these data are derived directly from a luminosity function that includes between 100 and 
450 stars in each 0.25~magnitude bin.  The SMC population has $-$1.4 $<$ [Fe/H] $<$ $-$1.0, and is therefore an 
order of magnitude more {\it metal-poor} than solar neighborhood studies, yet an order of magnitude more 
{\it metal-rich} than the ultra faint dwarf spheroidals.  Two recent projects have measured the shape of 
the stellar luminosity function in each of these regions, and for stellar populations where dynamical 
corrections are not needed.  First, at the metal-rich end, the IMF of the Galactic bulge was measured by 
\cite{zoccali00} using {\it HST}/NICMOS observations.  Although their overall best fit for a single slope 
is $\alpha$ = $-$1.33 $\pm$ 0.07, this slope is affected by a turnover at the lowest masses.  They point 
out that a two-slope IMF gives a better fit and has $\alpha$ = $-$2.00 $\pm$ 0.23 at $M >$ 0.5~$M_\odot$.   
Over the same mass range, we can also compare the luminosity function of the SMC (Table~1) to that of 
the Ursa Minor dwarf spheroidal galaxy which was studied by \cite{feltzing99} using {\it HST}/WFPC2.  
The distance to this galaxy is the same as the SMC, ($m - M$)$_0$ = 19.1 $\pm$ 0.1, and the stellar population 
is both very old and metal-poor ([Fe/H] = $-$2).  \cite{feltzing99} tabulate the star counts 
of this satellite, accounting for incompleteness, and demonstrate that the true luminosity function rises 
by a factor of 3.1 from $F606W$ = 23 to 27 (i.e., above the 50\% completeness limit).  Over the same luminosity 
range, the SMC luminosity function in our data rises by a factor of 2.8 and is therefore in excellent 
agreement with the dwarf spheroidal results.  Taken together, these studies demonstrate that the IMF is 
very similar in three unevolved and unrelaxed stellar populations with a large difference in metallicity of 
[Fe/H] = $>$0 to $-$2.0.

\section{Summary}

Knowledge of the mass distribution of stars in stellar populations such as clusters and galaxies is a 
fundamental input to a wide range of problems in astrophysics.  We have presented a new study of the 
IMF by analyzing high-precision {\it HST}/ACS photometry in one of the deepest images ever obtained for 
a nearby stellar population, the SMC.  We isolate the SMC main sequence from the CMD and resolve a 
high-precision luminosity function of the galaxy down to 29th magnitude.  This population is 
modeled by convolving input power-law IMFs with binaries, photometric scatter, and incompleteness.  
The best-fit IMF that reproduces the SMC population is $dN/dM \propto$ $M^{-1.90^{+0.15}_{-0.10}}$ 
(3$\sigma$ error), shallower than the Salpeter slope of $\alpha$ = $-$2.35.  We demonstrate that a single power law 
reproduces the completely mass distribution of stars from $M$ = 0.37 -- 0.93~$M_\odot$, suggesting 
that the stellar mass function does not break in this range.  At even lower masses, the data demonstrate a 
drop off in number counts that is inconsistent with an extrapolation of this IMF.  A comparison of the 
SMC IMF at $-$1.4 $<$ [Fe/H] $<$ $-$1.0 to that of very metal-rich stars in the Galactic bulge, as well 
as very metal-poor stars in the Ursa Minor dwarf spheroidal galaxy, indicates a similar slope and 
therefore a negligible metallicity gradient.

\acknowledgments

We are grateful to Tom Brown for several useful discussions related to constructing artificial star 
tests and for simulating CMDs of stellar populations.  We also thank Elena Sabbi for her help in providing 
background information on the SMC.  Support for program GO-11677 was provided by NASA through a 
grant from the Space Telescope Science Institute, which is operated by the Association of Universities 
for Research in Astronomy, Inc., under NASA contract NAS 5-26555.  HBR is supported by grants from The 
Natural Sciences and Engineering Research Council of Canada and by the University of British Columbia. \\




\noindent {\it Facilities:} \facility{HST (ACS)}


\clearpage

\end{document}